\documentclass[11pt, a4paper]{article}

\usepackage[margin=1in]{geometry}
\usepackage{amsmath}
\usepackage{amssymb}
\usepackage{graphicx}
\usepackage{times}
\usepackage{hyperref}
\hypersetup{
    colorlinks=true,
    linkcolor=blue,
    filecolor=magenta,      
    urlcolor=cyan,
}

\title{\textbf{The Metaphysics of Protection: Emergence, Agency, and the Ontological Status of Logical Qubits}}
\author{Kamran Majid}
\date{\today}

\begin{document}

\maketitle

\begin{abstract}
\noindent This paper argues that the practice of fault-tolerant quantum computation, specifically the mechanism of Quantum Error Correction (QEC), offers a profoundly new lens through which to examine foundational questions of ontology, emergence, and interpretation. We move beyond the standard debate on quantum speedup to ask: What is the nature of the entity---the logical qubit---that is being protected, and what does the active, goal-directed process of its protection reveal about physical reality? We argue that the logical qubit presents a unique case study in the metaphysics of identity, functioning as a quantifiable "Ship of Theseus" in Hilbert space. We introduce the concept of "engineered emergence" to describe the active, information-driven stabilization of the logical qubit, distinguishing it from passive forms of emergence and positioning it as a new category of causal structure. Finally, we demonstrate that the logical qubit serves as a powerful new testbed for major interpretations of quantum mechanics (including agent-centered, Many-Worlds, and Bohmian views), revealing novel strengths and challenges for each. We conclude that the technological imperative of fault-tolerance is not merely an engineering problem but a catalyst for deep philosophical insight, transforming abstract debates into concrete physical questions.
\end{abstract}


\hrulefill

\section{Introduction}

The philosophical discourse surrounding quantum computing has, for decades, been dominated by a single, captivating question: what is the source of its power? This "quantum speedup debate" has explored the roles of superposition, entanglement, and contextuality in enabling quantum algorithms to, in principle, solve certain problems exponentially faster than their classical counterparts \cite{cuffaro2012kantian, sergioli2018quantum, jozsa2003role}. While this line of inquiry is foundational, it has inadvertently overshadowed a more immediate and, arguably, more metaphysically revealing question. The future of quantum computing does not hinge on a final verdict in the speedup debate, but on a practical, technological necessity: fault-tolerance \cite{nielsen2010quantum, gottesman2009introduction, terhal2015quantum}. Without the ability to correct for the incessant errors to which quantum systems are prone, large-scale quantum computation remains a theoretical fantasy \cite{lidar2013quantum, aharonov1997fault}.

This report pivots the philosophical focus from the \textit{why} of quantum speed to the \textit{what} of quantum stability. It proposes a novel research program centered on the physical and metaphysical implications of Quantum Error Correction (QEC), the set of techniques designed to achieve fault-tolerance. The central contention is that QEC is not a mere engineering "fix" or a peripheral technical detail that philosophers can afford to ignore. Rather, it is a profound physical process that actively constructs and maintains the very entities that perform quantum computations. In doing so, it forces a direct confrontation with foundational issues of ontology, causality, and identity in a new, technologically-grounded context where abstract philosophical problems acquire concrete, physical manifestations \cite{harlow2016jerusalem, preskill1998fault}.

The protagonist of this investigation is the "logical qubit"---the robust, information-bearing unit encoded across a redundant collective of fragile "physical qubits." This research program moves beyond asking "Why is a quantum computer fast?" to ask a more fundamental set of questions:

\begin{enumerate}
    \item \textbf{What is the ontological status of a logical qubit?} Is it a real, emergent physical entity with its own distinct properties, or a mere calculational fiction? How does it maintain its identity through the constant flux of error and correction?
    \item \textbf{Does QEC represent a new kind of emergence in physics?} The stability of a logical qubit is not a natural, passive property but one that is actively engineered and maintained by a goal-directed, algorithmic process. Does this "engineered emergence" challenge our standard philosophical accounts of how levels of reality relate to one another?
    \item \textbf{How does the active intervention of QEC inform interpretations of quantum mechanics?} The continuous, subtle dialogue between a classical control system and a quantum system in QEC provides a novel testbed for all major interpretations, from agent-centered views like QBism to realist accounts like the Many-Worlds Interpretation and Bohmian Mechanics.
\end{enumerate}

By pursuing these interlocking questions, this report aims to demonstrate that the technological imperative of fault-tolerance is a powerful catalyst for philosophical progress. It connects disparate fields---quantum information theory, condensed matter concepts, metaphysics, and quantum foundations---in a synthesis that promises to transform long-standing debates. This approach is designed to be foundational in its aims, technically grounded in its methods, and philosophically novel in its conclusions, aligning with the rigorous standards of contemporary philosophy of physics \cite{wallace2022philosophy}.

\section{Part I: The Physical Foundation: From Fragile Matter to Robust Information}

To appreciate the philosophical depth of the logical qubit, one must first grasp the profound physical challenge it is designed to solve. The universe is hostile to quantum information. The very properties that grant quantum computers their potential power---superposition and entanglement---are exquisitely fragile. This section establishes the physical basis of Quantum Error Correction (QEC), framing the logical qubit as the actively constructed solution to the problem of decoherence. It provides the essential technical premises upon which the subsequent philosophical analysis will be built.

\subsection{The Imperative of Protection: Decoherence and the Quantum-to-Classical Transition}

The primary obstacle to building a functional, large-scale quantum computer is a ubiquitous physical process known as decoherence \cite{zurek2003decoherence, schlosshauer2007decoherence}. Decoherence describes the process by which a quantum system, initially in a pure state of superposition, loses its uniquely "quantum" characteristics due to its inevitable interaction with the surrounding environment. This interaction effectively entangles the quantum system with the countless degrees of freedom of its environment---air molecules, thermal photons, electromagnetic fields---in a way that rapidly degrades the delicate phase relationships that define the superposition \cite{zurek2003decoherence}.

From a foundational perspective, decoherence is a key mechanism in explaining the quantum-to-classical transition \cite{zurek2022quantum, wallace2012emergent}. It illuminates how the seemingly definite, classical world of our experience can emerge from an underlying quantum reality without needing to postulate a fundamental, mysterious "collapse" of the wavefunction. The environment acts as a vast, incessant observer, continuously "measuring" the quantum system and selecting a preferred basis (the "pointer basis"), causing the off-diagonal terms of the system's density matrix, which represent quantum coherence, to decay exponentially fast. The result is a system that, for all practical purposes, behaves classically.

While decoherence helps explain the emergence of the classical world, it is the principal antagonist in the narrative of quantum computation. The very entanglement with the environment that stabilizes classical reality is what destroys the fragile quantum states needed for computation \cite{nielsen2010quantum, lidar2013quantum}. An uncontrolled interaction of even a single environmental particle with a qubit can be enough to corrupt its state, introducing an error and derailing a complex calculation. The challenge is so immense that for a time, many prominent quantum theorists doubted that QEC was even possible in principle, believing that large-scale quantum computation would be forever forbidden by nature's insistence on erasing quantum coherence \cite{lidar2013quantum, zurek2003decoherence}. The imperative of protection, therefore, is absolute. Building a quantum computer is not merely about fabricating qubits; it is about engaging in an active, continuous battle against the natural tendency of the universe to "find out" what state those qubits are in and, in doing so, to destroy the computation.

A deeper examination reveals a subtle dialectic at play. QEC is not simply about shielding a system from decoherence; it is about mastering it. The core mechanism of error detection in most QEC schemes is the \textit{syndrome measurement} \cite{kitaev2003fault, fowler2012surface}. In this process, an auxiliary qubit, or "ancilla," is intentionally entangled with a specific subset of the data-carrying physical qubits. This ancilla is then measured. This procedure is carefully designed so that the measurement outcome reveals information about potential errors on the data qubits---the "error syndrome"---without revealing any information about the logical state itself, thereby avoiding the collapse of the computation.

This process of controlled entanglement and subsequent measurement of the ancilla is, in fact, a localized and highly engineered form of decoherence. The quantum computer deliberately "leaks" a specific, chosen piece of information (the error syndrome) into a controlled part of its environment (the measurement apparatus). This controlled information leak is then used to diagnose and reverse the effects of the uncontrolled, chaotic information leak to the wider, unmonitored environment. Therefore, QEC should not be understood as a simple negation of decoherence, but as a sophisticated harnessing of it. It represents a masterful manipulation of the physics of the quantum-to-classical transition, using the tools of measurement and entanglement to actively carve out and maintain a protected quantum subspace within a relentlessly classical world. This reframes QEC from a brute-force shield into an elegant, judo-like maneuver that turns the power of the environment against itself to achieve quantum stability.

\subsection{The Architecture of a Logical Qubit: Stabilizers and the Surface Code}

The solution to the problem of decoherence lies in a principle familiar from classical information theory: redundancy. To protect information from noise, one encodes it across a larger number of physical carriers. Quantum Error Correction adapts this principle to the quantum realm, encoding the information of a single "logical qubit" into a complex, entangled state of many "physical qubits" \cite{preskill1998fault, dennis2002topological}. The precise set of instructions for this encoding and its subsequent protection is known as a quantum error-correcting code. Such a code defines a small, protected subspace---the \textit{code space}---within the vast Hilbert space of the collective physical qubits \cite{lidar2013quantum, kitaev2003fault, gottesman1997stabilizer}.

The most prevalent and well-understood family of QEC codes are the \textit{stabilizer codes} \cite{kitaev2003fault, knill1997theory}. These codes have a particularly elegant mathematical structure that makes them amenable to analysis and implementation. The code space of a stabilizer code is defined as the simultaneous +1 eigenspace of a set of commuting operators, which are products of Pauli matrices ($X, Y, Z$). These operators are known as the \textit{stabilizer generators} \cite{gottesman1997stabilizer}. Any state $|\psi_L\rangle$ within the code space is, by definition, an eigenstate of every stabilizer generator $S_i$ with eigenvalue +1, such that $S_i |\psi_L\rangle = |\psi_L\rangle$.

The power of this formalism becomes apparent when considering errors. A common error, such as a bit-flip ($X$ error) or a phase-flip ($Z$ error) on a single physical qubit, will cause the state to anticommute with some of the stabilizer generators. Consequently, the errored state is no longer a +1 eigenstate of those generators; it becomes a -1 eigenstate instead. By measuring the eigenvalues of all the stabilizer generators, one can determine which of them have flipped from +1 to -1. This pattern of outcomes is the \textit{error syndrome}. Crucially, because the stabilizer operators all commute with the logical operators that define the encoded qubit state, this syndrome measurement reveals which error occurred without disturbing the logical information itself \cite{kitaev2003fault, fowler2012surface}. This is the central insight of QEC: it is possible to gain information about the errors without gaining any information about the state being protected. Once the syndrome is known, a classical computer acting as a controller can deduce the most likely error and apply a corresponding correction operation to return the system to the code space.

A paradigmatic and experimentally promising example of a stabilizer code is the \textit{surface code} \cite{fowler2012surface, dennis2002topological}. The surface code is a type of \textit{topological code}, where the protection of information derives from the global, topological properties of the system. In its most common implementation, physical qubits are arranged on the edges of a two-dimensional square lattice \cite{fowler2012surface}. There are two types of stabilizer generators: \textit{plaquette operators}, which are products of $Z$ operators around a face of the lattice, and \textit{star} or \textit{vertex operators}, which are products of $X$ operators around a vertex. One can verify that all these operators commute with each other, defining a valid stabilizer group.

Errors in the surface code have a beautiful physical interpretation. A single-qubit $X$ error will violate the two adjacent plaquette stabilizers, creating a pair of "excitations" on those plaquettes. Similarly, a $Z$ error creates excitations on the adjacent vertices. These excitations behave like particles, often called "anyons," and the error syndrome measurement is equivalent to detecting their locations \cite{fowler2012surface}. An error chain---a string of errors on multiple qubits---results in a pair of anyons at the endpoints of the string. The task of the classical controller, known as the \textit{decoder}, is to look at the pattern of anyons revealed by the syndrome measurements and infer the most probable error string that could have created them. It then applies a corresponding string of correction operators to annihilate the anyons and restore the state. The surface code is a leading candidate for building fault-tolerant quantum computers because its 2D layout with only nearest-neighbor interactions is compatible with current hardware fabrication techniques, and it possesses a relatively high \textit{error threshold}---the maximum physical error rate below which the code can reliably correct errors \cite{nielsen2010quantum, preskill1998fault}.

\subsection{The Non-Local and Collective Nature of Protected Information}

The robustness of a logical qubit stems directly from the way it stores information. Unlike a physical qubit, where the quantum state $|\psi\rangle = \alpha|0\rangle + \beta|1\rangle$ is a localized property of a single physical system, a logical qubit's information is profoundly non-local. It is not stored in any single physical qubit or any small subset of them. Instead, it is encoded in the intricate, long-range patterns of quantum entanglement that correlate the entire collective of physical qubits \cite{harlow2016jerusalem, preskill1998fault}.

This delocalization is most clearly seen by examining the \textit{logical operators}. These are the operators that act as the Pauli $X_L$ and $Z_L$ gates for the encoded logical qubit, allowing it to be manipulated for computation. In a stabilizer code like the surface code, these logical operators are themselves non-local. For a toric code (a surface code on a torus, a closed surface), the logical operators correspond to strings of single-qubit Pauli operators that wrap around the torus along its non-contractible loops \cite{fowler2012surface}. For a planar surface code with boundaries, they are strings connecting the different types of boundaries.

Because the logical operators are global and the stabilizer generators are local, any single, local error on a physical qubit will commute with the logical operators. This means a local error cannot change the logical state. To cause a \textit{logical error}---that is, to flip the encoded qubit from, for example, logical $|0_L\rangle$ to logical $|1_L\rangle$---an error must be applied that is itself a non-local operator of the same topological type as a logical operator. This requires a correlated string of errors spanning the entire code, an event whose probability decreases exponentially with the size of the code \cite{dennis2002topological}. The \textit{distance} ($d$) of the code is the minimum number of single-qubit errors that can produce a logical error. The larger the distance, the more robust the logical qubit. This is the essence of topological protection: the information is stored not in the local details of the physical qubits, but in the global, topological structure of their collective entangled state, making it immune to local perturbations \cite{fowler2012surface}.

The transformation from a fragile physical qubit to a robust logical qubit is so profound that it warrants a direct comparison of their fundamental properties. The following table summarizes the key distinctions that form the basis for the subsequent philosophical investigation.

\begin{table}[h!]
\centering
\caption{A Comparative Ontology of Physical and Logical Qubits}
\label{tab:qubit_comparison}
\begin{tabular}{|p{0.2\textwidth}|p{0.35\textwidth}|p{0.35\textwidth}|}
\hline
\textbf{Property} & \textbf{Physical Qubit} & \textbf{Logical Qubit (e.g., in a Surface Code)} \\
\hline
\textbf{Physical Substrate} & A single two-level quantum system (e.g., atom, superconducting circuit). & An entangled collective of many physical qubits arranged on a lattice. \\
\hline
\textbf{Information Storage} & State vector $\alpha|0\rangle + \beta|1\rangle$ localized on a single particle/system. & A non-local, entangled state in a protected subspace of the joint Hilbert space. \\
\hline
\textbf{Stability / Lifetime} & Fragile; subject to rapid decoherence from local environmental noise. & Robust; protected from local errors. Lifetime increases with the number of physical qubits (code distance). \\
\hline
\textbf{Identity Condition} & Persistence of the physical system itself. & Functional: persistence of the ability to store information with a low logical error rate, maintained by active correction. \\
\hline
\textbf{Interaction with Env.} & Any local interaction can corrupt the state (decoherence). & Local interactions create detectable, correctable error syndromes without corrupting the logical state. \\
\hline
\end{tabular}
\end{table}

\section{Part II: The Ontological Status of the Logical Qubit}

Having established the physical architecture of the logical qubit, we can now turn to the first of our core research questions: What \textit{is} this entity? Is it a genuine physical object, a mere calculational convenience, or something else entirely? To answer this, we must bring the tools of metaphysics---specifically, the study of identity, persistence, and ontology---to bear on the concrete physical system described in Part I. The logical qubit, we will find, serves as a remarkable test case that pushes classical metaphysical problems into a new, quantifiable domain.

\subsection{Persistence and Identity: The Ship of Theseus in Hilbert Space}

One of the most venerable problems in metaphysics is the puzzle of identity through time, famously illustrated by the Ship of Theseus paradox \cite{lewis1986plurality, sider2001four}. If the planks of Theseus's ship are replaced one by one until no original planks remain, is it still the same ship? This thought experiment probes the conditions under which an object can persist despite a complete change in its constituent matter. It forces a distinction between an object's material composition and its form or identity.

The logical qubit provides a precise and physically realized analogue of this ancient puzzle. It is a veritable Ship of Theseus navigating the waters of Hilbert space. The "planks" of the logical qubit are the quantum states of its constituent physical qubits. These planks are not merely replaced over time; they are in a state of \textit{constant, deliberate flux}. The QEC cycle is a dynamic process of continuous monitoring and intervention. At each step, syndrome measurements are performed, and based on their outcomes, corrective operations are applied to the physical qubits \cite{kitaev2003fault, fowler2012surface}. This means that the underlying quantum state of the many-qubit ensemble is perpetually changing. The system is actively "repaired" from one moment to the next.

The logical qubit persists not because its physical substrate is static---it is anything but---but because its \textit{information-bearing structure} is actively and continuously restored to the protected code space. Its identity is not one of material stasis but of functional stability. This dynamic existence forces us to confront classical metaphysical theories of persistence. An \textit{endurantist} view holds that an object is "wholly present" at every moment of its existence \cite{lewis1986plurality}. It is difficult to see how a logical qubit could be wholly present when its underlying physical description is constantly being rewritten. A \textit{perdurantist} view, which conceives of objects as four-dimensional "spacetime worms" composed of temporal parts, seems more apt \cite{lewis1986plurality}. On this account, the logical qubit would be the entire four-dimensional process, and its "persistence" would be the continuity of this process through time, with each post-correction state serving as a temporal part. The active, process-based nature of its being strongly favors a metaphysical account that prioritizes process and function over static substance.

This connection does more than simply provide a new example for an old puzzle. It offers a way to sharpen the puzzle itself. Traditional metaphysical debates about persistence often founder on the vagueness of the identity conditions. When, precisely, does the ship cease to be the same ship? The question seems to lack a non-arbitrary answer \cite{sider2001four}. The logical qubit, however, introduces a remarkable level of precision into the discussion. Its identity is inextricably linked to its function: the reliable storage of quantum information. This function is not a vague, qualitative property but a precisely quantifiable one, measured by the \textit{logical error rate}---the probability that the stored information becomes corrupted despite the correction process \cite{preskill1998fault, kitaev2003fault}.

The very foundation of fault-tolerant quantum computation rests on the \textit{threshold theorem}. This theorem states that if the error rate of the individual physical components is below a certain critical value (the threshold), then the logical error rate can be made arbitrarily low simply by increasing the number of physical qubits used in the code (i.e., increasing the code's distance) \cite{aharonov1997fault, knill1997theory}. This provides a clear, non-arbitrary criterion for the persistence of the logical qubit. The logical qubit \textit{is} a logical qubit---it maintains its identity---if and only if the QEC protocol successfully keeps its logical error rate below a threshold required for a given computation to succeed. Its existence is not a simple binary property (it is or it isn't) but a graded one, tied directly to a measurable performance metric.

This transforms the qualitative philosophical puzzle of identity into a quantitative physical question. It suggests a new, functionalist criterion for the identity of certain complex systems: an object persists as long as it successfully performs its defining function above a specified, non-arbitrary performance threshold. The logical qubit thus offers a path to operationalize and potentially resolve certain metaphysical debates by grounding them in the concrete realities of physics and information theory.

\subsection{The Quasiparticle Analogy: Real Entity or Useful Fiction?}

To better situate the logical qubit within the landscape of physics, it is useful to search for an ontological analogue. The most promising candidate comes from the field of condensed matter physics: the \textit{quasiparticle} \cite{anderson1963plasmons}. A quasiparticle is an emergent phenomenon in a complex many-body system, where a collective excitation behaves in many ways as if it were a single, fundamental particle. Examples include the phonon (a quantum of lattice vibration), the exciton (a bound electron-hole pair), and the magnon (a quantum of a spin wave) \cite{anderson1963plasmons}.

The philosophical status of quasiparticles has long been a subject of debate \cite{castellani2002meaning}. Are they "real" physical entities, or are they merely a convenient mathematical fiction that simplifies the intractable problem of describing the interactions of $\sim10^{23}$ particles? \cite{anderson1963plasmons}. On one hand, they are emergent, existing only within the medium of the solid and not in a vacuum. On the other hand, they have well-defined properties like mass and charge (often different from their constituent particles) and can be experimentally observed to scatter and interfere.

The logical qubit shares many of the defining characteristics of a quasiparticle. It is a collective phenomenon, arising from the interactions of many constituent parts (the physical qubits). Its most important property---robustness---is a property of the collective, not of any individual constituent. And it can only "exist" within the specific context of the larger system (the quantum computer operating under a QEC protocol). The logical qubit, like a quasiparticle, is a way of describing the simple, effective behavior of a complex, many-body quantum system.

However, the analogy, while illuminating, ultimately breaks down at a crucial point, revealing the unique nature of the logical qubit. A typical quasiparticle, like a phonon or an electron polaron, arises from the \textit{natural, passive dynamics} of the system. Its existence and properties are determined by the system's Hamiltonian and the intrinsic interactions among its constituent particles. It is a natural consequence of the physics of the collective.

The logical qubit is fundamentally different. Its defining property---its enhanced stability against decoherence---is not a natural property of the ensemble of physical qubits. In fact, the natural tendency of the system is to decohere and lose the encoded information. The stability of the logical qubit is the result of \textit{active, external, and goal-directed intervention} \cite{preskill1998fault, google2023suppressing}. It is maintained by a classical control system that is continuously performing measurements, processing information via a specific algorithm (the decoder), and applying corrective feedback. The logical qubit does not simply emerge; it is \textit{engineered}. This points to the need for a new ontological category, one that can account for this active, algorithmic maintenance.

\subsection{A New Ontological Kind? Information Made Manifest}

Synthesizing these analyses, the logical qubit appears to resist easy categorization within our existing ontological frameworks. It is not a fundamental particle. It is not a standard substance in the metaphysical sense. It is analogous to a quasiparticle but distinguished by the active, engineered nature of its existence. To fully grasp its status, we can turn to the foundational debates within quantum mechanics itself, particularly the long-standing question of the reality of the quantum state.

This debate is often framed as a dichotomy between $\psi$-ontic and $\psi$-epistemic interpretations \cite{harrigan2010einstein}. In $\psi$-ontic models, the wavefunction ($\psi$) corresponds directly to a state of physical reality; different wavefunctions correspond to different realities. In $\psi$-epistemic models, the wavefunction represents a state of knowledge or belief about an underlying, potentially hidden, reality; different wavefunctions could correspond to the same underlying reality \cite{harrigan2010einstein, pusey2012reality}.

The logical qubit complicates this picture in a fascinating way. On one hand, its state is encoded in a definite, physical, many-body wavefunction, which evolves according to the Schrödinger equation (between corrections). This suggests a $\psi$-ontic character. On the other hand, its very identity and persistence are defined by its function as an information-bearer, and its stability is a measure of how well that information is preserved against error. This informational character has a distinctly $\psi$-epistemic flavor. The logical qubit seems to be both physical reality and information at the same time.

This suggests that the logical qubit may represent a new ontological kind, one that transcends the traditional $\psi$-ontic/$\psi$-epistemic divide. It is not simply a physical system, nor is it simply information. It is an \textit{information-theoretic structure made physically robust through active, dynamical engineering}. Its essence is informational---to store one qubit of data---but its existence is a concrete, physical, and dynamical achievement. This moves the philosophical debate beyond simply asking whether the quantum state is "real" and toward a more nuanced analysis of how \textit{information itself} can be stabilized and reified as a persistent physical entity. This entity, which we can term a \textit{stabilized information-bearer}, connects abstract discussions in the philosophy of quantum information \cite{sergioli2018quantum, timpson2013quantum, bub2016bananaworld} with concrete questions of physical realism \cite{maudlin2019philosophy, wallace2020plurality}. The logical qubit is what information looks like when it is forced to survive in the physical world.

\section{Part III: Engineered Emergence: A New Causal Structure in Physics}

The unique ontological status of the logical qubit, characterized by its active, goal-directed maintenance, suggests that the process by which it comes into being may also be unique. This part of the report develops our second core research question, arguing that Quantum Error Correction exemplifies a novel form of emergence that is distinct from the categories traditionally discussed in the philosophy of science. We term this "engineered emergence" and explore its unique causal structure, arguing that it represents a significant new concept for understanding the relationship between different levels of reality, particularly in the context of advanced technology.

\subsection{Beyond Passive Emergence: The Active Nature of QEC}

The philosophical literature on emergence typically distinguishes between two main types: weak and strong emergence \cite{bedau2008emergence, kim1999making}. \textit{Weak emergence} describes properties that, while novel at a macroscopic level, are in principle deducible from the dynamics of the system's microscopic constituents. The emergent property is a predictable, albeit complex, consequence of the lower-level laws. Standard examples from physics, such as the emergence of temperature and pressure in a gas from the statistical mechanics of molecular motion, fall into this category \cite{kim1999making}. These are cases of \textit{passive emergence}: the higher-level properties arise from the undirected, statistical behavior of the parts. They are a byproduct of the system's natural evolution.

\textit{Strong emergence}, by contrast, posits the existence of higher-level properties that are genuinely irreducible to the lower levels and possess novel, downward causal powers \cite{bedau2008emergence}. Such properties would not be predictable even with complete knowledge of the micro-physics. Consciousness is often proposed as a candidate for strong emergence, though its existence in nature is highly controversial, with some critics deeming it "uncomfortably like magic" \cite{bedau2008emergence}.

The emergence of the logical qubit's stability fits neatly into neither category. It is unlike passive, weak emergence because its defining property---robustness---is not a natural or statistically inevitable consequence of the interactions between the physical qubits. Left to their own devices, the qubits would simply decohere. The stability is actively \textit{imposed} upon the system from the outside. Yet, it is not strong emergence, as there is no appeal to irreducible new laws of nature. The entire process is, in principle, describable by the known laws of quantum and classical mechanics.

We therefore propose a third category: \textit{engineered emergence}. Engineered emergence describes the creation of a robust, stable, higher-level property through a goal-directed, information-processing feedback loop. The key features are:
\begin{enumerate}
    \item \textbf{Active Maintenance:} The emergent property is not a static or passive consequence of the substrate's nature but is actively and continuously maintained by an external process.
    \item \textbf{Goal-Directedness:} The process is governed by an algorithm designed to achieve a specific, pre-defined goal---in this case, the preservation of a logical quantum state with a low error rate \cite{preskill1998fault, google2023suppressing}.
    \item \textbf{Information-Based Feedback:} The maintenance loop is informational. The state of the lower level (the physical qubits) is monitored to extract information (the error syndrome), which is then processed by a higher-level system (the classical controller) to determine a corrective action that is fed back to the lower level.
\end{enumerate}

This concept can be distinguished from related ideas like \textit{active matter}. In active matter systems, such as flocks of birds or bacterial colonies, individual constituents consume energy to create collective motion and novel phases of matter \cite{marchetti2013hydrodynamics, ramaswamy2010mechanics}. While this is a form of active emergence, the "control" is distributed and local to each agent. Engineered emergence, as seen in QEC, is a step beyond this. The control is centralized, algorithmic, and explicitly aimed at stabilizing an abstract informational property rather than just generating a physical pattern.

\subsection{The Causal Role of the Classical Controller: Algorithmic Downward Causation}

The causal structure of engineered emergence is particularly noteworthy. The relationship between the classical controller and the quantum substrate constitutes a clear feedback loop. The quantum system's state influences the classical controller's information state via the syndrome measurements. The controller's algorithm then dictates a physical operation to be applied back to the quantum system.

This structure can be interpreted as a concrete, physical realization of \textit{downward causation}. In philosophical debates, downward causation---the idea that higher levels of a system can exert causal influence on their lower-level constituents---is often problematic and difficult to defend against charges of causal overdetermination \cite{bedau2008emergence}. However, in the case of QEC, the causal chain is explicit and undeniable. The "goal" of the system, which exists only at the abstract, informational level of the classical controller's algorithm, directly causes changes in the micro-level dynamics of the individual physical qubits. The high-level purpose (preserving the logical state) determines the low-level physical evolution.

This challenges standard philosophical accounts of reductionism and the relationship between levels of reality \cite{bedau2008emergence, o2020emergent}. The causal powers that define the higher level (the logical qubit's stability) are not intrinsic to that level. Instead, they are conferred and continuously sustained by the computational logic of an external system. This is not the mysterious downward causation of strong emergence, but a transparent, \textit{algorithmic downward causation}.

This leads to a profound conclusion about the nature of the regularities we observe. A law of physics is typically understood as an objective, mind-independent regularity in the behavior of physical systems. The stability of a logical qubit is a highly regular and predictable behavior. It follows a "law": local errors on its constituent parts will be detected and corrected, and the logical information will persist with high fidelity. This observed regularity, however, is not derived from the Schrödinger equation alone. It is the result of the Schrödinger equation governing the physical qubits \textit{plus} the deterministic execution of the QEC code's decoding algorithm by the classical controller \cite{preskill1998fault, fowler2012surface}.

The QEC process thus represents the \textit{physical instantiation of an algorithm as a law-like regularity}. The abstract, mathematical logic of the error-correcting code becomes a concrete, causal factor shaping the behavior of a physical system. This blurs the traditional, sharp distinction between a physical law and a computational algorithm. It suggests that, through technology, we can engineer new, effective physical laws that govern the behavior of emergent systems. This has significant implications for the philosophy of technology and the philosophy of science, hinting that the set of "laws" governing the phenomena in our world might not be fixed and immutable, but can be actively expanded by the intervention of intelligent, goal-directed agents.

\subsection{Spacetime as a Quantum Error-Correcting Code: The Ultimate Emergence}

The concept of engineered emergence, born from the technological project of quantum computing, may have echoes in the most fundamental structures of the cosmos. A speculative but powerful line of research in quantum gravity has uncovered a remarkable connection between the emergence of spacetime and the principles of quantum error correction \cite{almheiri2015bulk, pastawski2015holographic}.

In the context of the anti-de Sitter/conformal field theory (AdS/CFT) correspondence, a leading model in quantum gravity, spacetime itself is not fundamental. The robust, classical geometry of the higher-dimensional "bulk" spacetime is understood to emerge holographically from the quantum dynamics of a highly entangled system of particles living on its lower-dimensional boundary \cite{almheiri2015bulk}. In 2014, physicists Ahmed Almheiri, Xi Dong, and Daniel Harlow discovered that this holographic emergence works precisely like a QEC code \cite{almheiri2015bulk}. The information about a point deep within the bulk spacetime is encoded redundantly across the boundary. This information can be reconstructed from different, overlapping patches of the boundary, just as a logical qubit's state can be reconstructed from different subsets of its physical qubits. The intrinsic robustness of spacetime against local quantum fluctuations, on this view, is a direct consequence of the error-correcting properties of the underlying entanglement structure of the boundary theory \cite{harlow2018tasi}.

This "It from Qubit" research program suggests that the very fabric of reality might be a form of emergent, error-corrected structure. The stability of our macroscopic world could be a manifestation of a cosmic QEC protocol that protects the universe's quantum information. While this remains a highly theoretical and speculative area of research, it provides a profound example of the potential scope and significance of the principles at play. It suggests that the engineered emergence we are beginning to construct in our laboratories might be a toy model of the universe's own fundamental construction principle. What we learn from building a quantum computer may, in the end, teach us about the nature of space and time itself.

To clarify the novelty and significance of engineered emergence, it is useful to place it within a taxonomy of different types of emergence.

\begin{table}[h!]
\centering
\caption{A Taxonomy of Emergence}
\label{tab:emergence_taxonomy}
\begin{tabular}{|p{0.25\textwidth}|p{0.3\textwidth}|p{0.2\textwidth}|p{0.2\textwidth}|}
\hline
\textbf{Type of Emergence} & \textbf{Causal Mechanism} & \textbf{Locus of Control / Goal} & \textbf{Key Example} \\
\hline
\textbf{Weak Emergence} & Statistical aggregation of lower-level dynamics. & None (undirected). The emergent property is a passive byproduct. & Temperature/Pressure in a gas. \\
\hline
\textbf{Strong Emergence} & Postulated irreducible, novel causal powers at the higher level. & Internal to the system (if any). & Consciousness (hypothesized). \\
\hline
\textbf{Engineered Emergence} & Active, information-based feedback loop between levels. & External. The emergent property is actively maintained to meet a pre-defined goal. & Stability of a Logical Qubit. \\
\hline
\end{tabular}
\end{table}

\section{Part IV: Re-evaluating Quantum Interpretations Through the Lens of Protection}

The logical qubit is not merely a new object for metaphysical analysis; it is a new tool for the philosophy of physics. As a concrete, complex, and well-understood physical system, it can serve as a novel testbed for the major interpretations of quantum mechanics. Each interpretation makes claims about the fundamental nature of reality, the status of the quantum state, and the role of measurement. By asking each interpretation to account for the existence and stability of a logical qubit, we can expose their underlying assumptions, reveal new challenges, and identify unique explanatory strengths. This final part of the report uses the logical qubit as a lens to probe and differentiate these foundational frameworks.

\subsection{The Agent's Dialogue: Implications for QBism and Relational QM}

Agent-centered interpretations of quantum mechanics shift the focus from a "view from nowhere" to the role of the observer or agent in constituting quantum reality. The QEC process, with its explicit dialogue between a classical controller and a quantum system, provides a rich and concrete physical model for the abstract claims of these interpretations.

\textbf{Quantum Bayesianism (QBism)} posits that a quantum state does not represent an objective property of a physical system, but rather an agent's personal degrees of belief about the likely outcomes of their future interactions with that system \cite{fuchs2013quantum}. A measurement is an action an agent takes on the world, and the outcome is the personal experience that updates their beliefs \cite{fuchs2013quantum, von2016qbism}. QBism reinterprets the Born rule as a normative constraint on rational belief formation.

The QEC process maps onto the QBist framework with remarkable fidelity. The classical controller can be seen as the physical proxy for the QBist "agent." The logical state it aims to protect corresponds to the agent's initial set of beliefs (or "priors") about the information encoded. The QEC cycle is then a perfect model of the agent's interaction with the world. The agent performs an action (a syndrome measurement) to acquire new information (the error syndrome). This new information leads to an update of the agent's beliefs about the state of the physical qubits, which in turn prompts a new action (a corrective operation). The goal of this entire process is to ensure that the agent's core belief---the logical state---remains a reliable guide for future actions (i.e., for the final readout of the computation). From this perspective, the logical qubit \textit{is} a stabilized belief. Its reality is not that of a static, mind-independent object, but is grounded in the successful, continuous, and physically real process of belief-updating performed by the agent's computational proxy.

\textbf{Relational Quantum Mechanics (RQM)}, developed by Carlo Rovelli, holds that the quantum state of a system is not an absolute property but is always relative to another system, the "observer" \cite{rovelli1996relational, laudisa2019relational}. A physical "event" occurs when two systems interact, causing a variable of one system to acquire a definite value \textit{relative to the other} \cite{di2022relational}. There is no absolute, universal state of affairs, only a network of relations.

Again, the logical qubit provides a powerful illustration. The logical state can be understood as existing \textit{relative to the classical controller}. The syndrome measurements are precisely the "events" of RQM, where the error properties of the physical qubits become definite relative to the controller's ancillas. The remarkable stability of the logical qubit is, in RQM's terms, the stability of this \textit{informational relation} between the quantum register and the classical controller. Recent developments in RQM have addressed the potential for solipsism by emphasizing that this relational information must be physically encoded and accessible to other observers, for instance through a "Cross-perspective links" postulate \cite{adlam2023information}. QEC provides a concrete mechanism for precisely this. The logical state, as a relation to the controller, is encoded in the robust, collective physical state of the qubit array. Its correctness can be verified by a second agent who interacts not with the fragile quantum system directly, but with the classical controller, which holds a stable physical record of the information.

This leads to a deeper point about these agent-centered views. A primary challenge they face is explaining intersubjective agreement: if quantum states are personal beliefs or relative facts, why do different scientists performing the same experiment consistently get the same results? \cite{rovelli1996relational, pienaar2021qbism}. The logical qubit offers a compelling model for how this can happen. The "state" of the logical qubit is not a private, disembodied belief. It is a state relative to a \textit{physical classical control system}. This system, by its nature, maintains a robust, classical record of the information it is protecting. A second agent can then interact with the \textit{first agent's controller} to verify the logical state. The QEC process, therefore, is a physical mechanism that transforms a "relative fact" or a "personal belief" into a robust, externalized, and physically encoded piece of information that is stable enough to be shared and agreed upon by a community of observers. It physicalizes the process of creating shared knowledge, offering a powerful, concrete response to the charge of solipsism that has often been leveled against these interpretations.

\subsection{The Realist's Challenge: Locating the Logical Qubit}

Realist interpretations of quantum mechanics posit an observer-independent reality, described by some fundamental ontology (or "beables"). These interpretations must provide a clear account of what in their ontology corresponds to the logical qubit. This strange, non-local, and actively maintained entity poses unique and revealing challenges for them.

\subsubsection{In the Universal Wavefunction (Many-Worlds Interpretation)}

The Many-Worlds Interpretation (MWI) asserts that the universal wavefunction is objectively real and evolves deterministically according to the Schrödinger equation at all times. There is no collapse. Instead, every quantum measurement causes the universe to branch into multiple "worlds," one for each possible outcome, all of which are equally real \cite{everett1957relative, dewitt1973many, vaidman2018many}.

How does QEC function within this vast, branching multiverse? When a random error occurs---say, a bit-flip on a single physical qubit---the universe presumably branches. One set of branches corresponds to the "error world" where the flip occurred, and another to the "correct world" where it did not. The QEC process must then be interpreted within this framework.

One possibility is that QEC is a mechanism for \textit{branch management}. The syndrome measurement, in this view, is an interaction that identifies which branch the system is on (the "error" branch). The subsequent correction operation is a unitary evolution that acts on that branch to make it, from the perspective of the logical information, identical to the "correct" branch. This could be seen as a form of branch pruning or re-merging, actively steering errant parts of the multiverse back into computational alignment.

A second, perhaps more elegant, possibility is that the logical qubit is a feature that is inherently robust \textit{across worlds}. The QEC protocol is designed such that, despite the myriad branches created by different random error processes, the \textit{logical information remains an invariant} across the vast majority of them \cite{dewitt1973many, vaidman2018many}. The logical qubit is then an emergent, stable property of a whole decohered sector of the universal wavefunction, a pattern that persists across a multitude of parallel worlds. Its robustness is a feature of the multiverse itself, engineered by the QEC code.

Both possibilities raise a crucial question for MWI: what is the ontological status of the classical decoder? This computational device must exist and function correctly in all relevant branches to interpret the syndromes and dictate the corrections. Its own classical robustness and deterministic functioning seem to be a necessary precondition for engineering the quantum robustness of the logical qubit across the multiverse. The MWI must therefore rely on the stable existence of classical objects to explain how stable quantum objects are constructed, introducing a potential circularity or at least a deep dependence between the classical and quantum levels within its ontology.

\subsubsection{Among the Beables (Bohmian Mechanics)}

Bohmian Mechanics (BM) is a realist, deterministic interpretation that posits a dual ontology: there is the universal wavefunction, which evolves according to the Schrödinger equation, and there is an actual configuration of particles with definite positions at all times (the "beables") \cite{bohm1952suggested, durr2009bohmian, goldstein2017bohmian}. The wavefunction's role is to "guide" the motion of the particles via a guiding equation. All measurements, fundamentally, are measurements of particle positions \cite{bricmont2016making, norsen2011theory}.

In the Bohmian picture, a logical qubit cannot be a fundamental particle. It must be a property of the \textit{guiding wave} for the entire, high-dimensional configuration space of the physical qubits and the measurement apparatus. The non-local information defining the logical state must be encoded in the complex structure of this guiding field.

The active process of QEC presents a significant challenge and a rich area of inquiry for BM. The "measurement" of an error syndrome is a physical interaction that alters the total wavefunction of the system plus apparatus. According to the Bohmian framework, this change in the wave must instantly and non-locally alter the guiding field for all particles involved in the computation, thereby changing their velocities. The subsequent correction operation is another intervention that again sculpts the wavefunction. The stability of the logical qubit is the remarkable result of this carefully choreographed, non-local dance of the guiding wave, which in turn directs the trajectories of the underlying particle configuration.

This raises profound questions for the Bohmian. How, precisely, is the abstract, non-local, and binary information of the logical state represented by the continuous, concrete configuration of particles (the beables)? Is the logical state encoded in the collective arrangement of the particle positions, or is it purely a feature of the wave, with the particles' configuration merely being a secondary manifestation? The active, algorithmic nature of QEC places enormous causal and computational responsibility on the wavefunction. It must not only guide the particles but must do so in a way that reflects the execution of a complex classical decoding algorithm. This puts severe strain on the Bohmian picture, demanding an account of how the abstract logic of the decoder is translated into the physical dynamics of the guiding wave and the resulting particle trajectories \cite{bricmont2016making, esfeld2014ontology}.

The following table provides a culminating summary of how the logical qubit is situated within each of these major interpretive frameworks, highlighting the unique ontological commitments and explanatory challenges it raises for each.

\begin{table}[h!]
\centering
\caption{The Ontological Status of the Logical Qubit Across Major Interpretations}
\label{tab:interpretations}
\begin{tabular}{|p{0.25\textwidth}|p{0.35\textwidth}|p{0.35\textwidth}|}
\hline
\textbf{Interpretation} & \textbf{Ontological Status of the Logical Qubit} & \textbf{Role of the QEC Process} \\
\hline
\textbf{QBism} & A robust, stable set of beliefs held by an agent (or their classical proxy) about the outcomes of future interactions. & The physical process of an agent acting on the world (syndrome measurement) to update beliefs (apply corrections) and maintain a consistent "handbook" for future actions. \\
\hline
\textbf{Relational QM} & A stable, physically encoded informational relationship between the quantum system and a classical observer system (the controller). & The set of physical interactions ("events") that establish and maintain the robust correlation between the observer and the system, making the relation intersubjectively accessible. \\
\hline
\textbf{Many-Worlds} & A stable property of the universal wavefunction across a vast set of decohered branches ("worlds"). & A mechanism that either prunes/re-merges branches or ensures the logical information is an invariant across the multiverse, thus defining a stable set of worlds. \\
\hline
\textbf{Bohmian Mechanics} & A complex, non-local, and robust feature of the guiding wave, which in turn directs the motion of the underlying particles (beables). & A series of interventions that actively sculpt the guiding wave to maintain its information-bearing structure, with profound implications for non-locality and the wave's causal role. \\
\hline
\end{tabular}
\end{table}

\section{Conclusion: The Philosophical Significance of Fault-Tolerance}

This report has argued for a fundamental shift in the philosophical investigation of quantum computing: a pivot from the question of computational power to the metaphysics of computational stability. The technological necessity of fault-tolerance, far from being a mere engineering footnote, opens up a rich and unexplored territory for the philosophy of physics. The central object in this new landscape, the logical qubit, is not just a component in a future machine but a philosophical object of profound importance.

Our analysis has shown that the logical qubit forces a re-examination of core metaphysical concepts. It presents a physically realized, quantifiable version of the Ship of Theseus puzzle, transforming the vague problem of identity through change into a precise, functional, and threshold-dependent property. In doing so, it suggests a new, operationalized way to think about the persistence of complex systems. Its unique mode of existence, resisting easy classification as either a fundamental particle or a standard emergent property, has led us to propose it as a new ontological kind: a \textit{stabilized information-bearer}, an entity whose essence is informational but whose existence is a physical, dynamical achievement.

Furthermore, the process by which the logical qubit is maintained, Quantum Error Correction, demands a new category in our understanding of causality and emergence. We have introduced and defended the concept of \textit{engineered emergence} to describe the active, goal-directed, and information-driven feedback loop that creates a robust higher-level entity from a fragile lower-level substrate. This form of algorithmic downward causation, where the abstract logic of a computer program becomes a law-like, causal factor in the physical world, blurs the line between computation and physical law and has deep implications for our understanding of the relationship between mind, matter, and technology.

Finally, the logical qubit serves as an unparalleled instrument for probing the foundations of quantum mechanics itself. It provides a concrete physical scenario that tests the commitments of every major interpretation. For agent-centered views like QBism and RQM, it offers a powerful model for the physical realization of belief and the establishment of intersubjective agreement. For realist interpretations like MWI and Bohmian Mechanics, it poses sharp and revealing challenges, forcing them to locate this strange, non-local, and actively maintained entity within their respective ontologies.

The central theme of this work is that the practical, physical problem of protecting quantum information is inseparable from our deepest philosophical questions about the nature of reality. The road to building a functional quantum computer runs directly through the heart of metaphysics. By taking the physics of fault-tolerance seriously, philosophers have an opportunity to move beyond long-standing debates and engage with new questions forged at the cutting edge of science and technology. The metaphysics of protection is not just about how we will build the computers of the future; it is about what they will teach us about the universe and our place within it.

\end{document}